# The timing precision of transit light-curves

H.J Deeg[1,2], M. Seidel[3], and the Corot Photometric Follow-Up Team

[1] *Instituto de Astrofísica de Canarias, Tenerife, Spain [hdeeg@iac.es]*, [2] *Universidad de La Laguna, Tenerife, Spain*, [3] *Jacobs University, Bremen, Germany*

**Abstract.** Reliable estimations of ephemeris errors are fundamental for the follow-up of CoRoT candidates. An equation for the precision of minimum times, originally developed for eclipsing binaries, has been optimized for CoRoT photometry and been used to calculate such errors. It may indicate expected timing precisions for transit events from CoRoT, as well as from Kepler. Prediction errors for transit events may also be used to calculate probabilities about observing entire or partial transits in any given span of observational coverage, leading to an improved reliability in deductions made from follow-up observations.

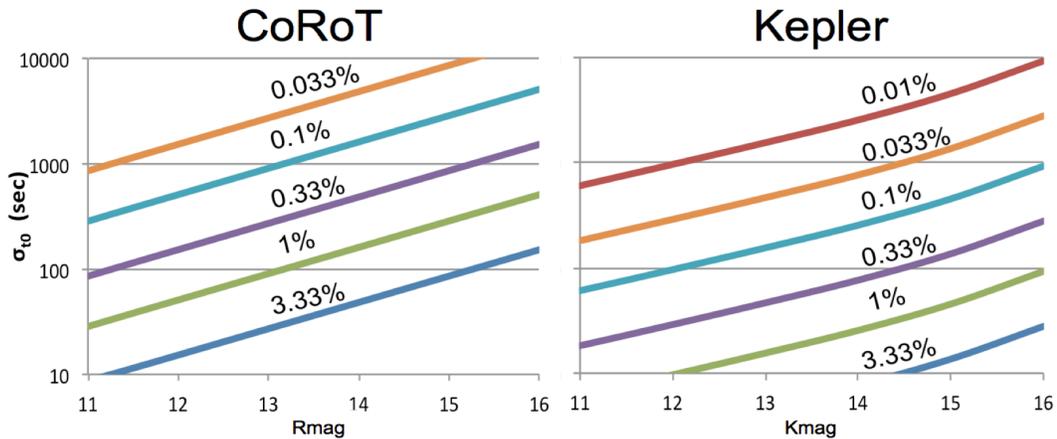

Figure 1.: *Timing precision of single transits with different amplitudes. Calculated for transits with 1 hr long ingress/egress and magnitude-dependent noises over times-scales of 1hr. Noises for CoRoT are scaled from noises over 2 hr given by Aigrain et al. (2009) and noises for Kepler are scaled from noises over 0.5 hr given by Jenkins et al. (2010) .*

The derivation of precise transit timings from light-curves is of relevance for several applications, mainly towards the detection of moons and of further orbiting planets. In the context of eclipsing binaries, timing information may also be indicative of physical changes within a binary system. Precise ephemeris are however also needed to predict times of transit or eclipse events for future observations, after the original coverage has ended. In the context of the CoRoT mission, the ground based photometric follow-up depends on the reliability of such predictions, but they are also important for observations of e.g. the Rossiter-McLaughlin effect. CoRoT observes stellar fields with a fairly large psf of ≈ 35"x23" and its photometry is obtained in large optimized apertures. A faint transit signal detected by CoRoT may therefore arise from either the target or from any other nearby source – typically eclipsing binaries in the background. Ground-based photometric follow-up has a much higher spatial resolution than the satellite and may hence detect these binaries. Its outcome can be either of the following:

i): the transit is reproduced on target ⇒ candidate verified ii): an eclipse is found on a nearby star⇒ false alarm

iii): no signal is found on either target or nearby star

⇒ the transit candidate is verified to be 'on-target' IF an on-target transit is too shallow for ground-based detection AND IF it can be demonstrated that none of the nearby stars has a signal strong enough to be a false alarm. This requires a calculation of the eclipse amplitude that these stars would need in order to mimick CoRoTs signal (Deeg et al. 2009) AND it requires it that follow-up observations were made during an eclipse event.

With iii) being the most common outcome, reliable predictions for the times of eclipse events become critical for a useful ground based follow-up. Estimations for the precision of a transit prediction can be derived from a transit-lightcurve's basic parameters (transit depth, duration, periodicity and the curve's length and it noise), based on an equation for the timing precision of single eclipses (Doyle & Deeg 2004). Fig. 1 shows this single-eclipse timing precision for typical transits expected from CoRoT and Kepler. This equation is currently used in predictions for CoRoT's ground-based follow-up; its reliability has been demonstrated from several cases were predicted eclipse timings could be compared to posterior timings found in follow-up observations.

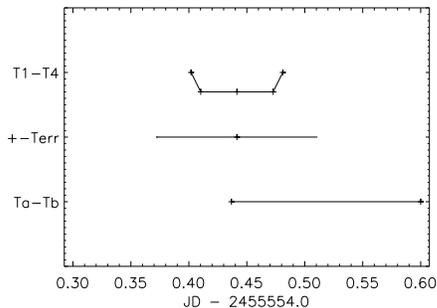

Figure 2.: *Predicted transit event and time of its observation: The upper line shows the predicted time of a transit whose sections are indicated scematically. The next line indicates the ± 1 sigma prediction error. The lowest line indicates the time-span when observations were performed.*

For the decision if a ground-based *non*-detection of eclipses on any contaminating star does imply an 'on-target' verdict (see case iii above), we need to estimate the likelihood that a ground-based observation has covered the eclipse event. Assuming a normal distribution of the prediction error $\sigma_t$, the probabilities that any of the eclipse contacts T1,..,T4, – or eclipse sections such as the ingress – are within the observational coverage can be derived from Error-Functions. Fig. 2 shows an example of a planet candidate which was detected by CoRoT in winter 2008/09 and for which follow-up observations were taken 2 years later, after a significant timing-error of $\pm 1.6h$ had accumulated: The probability to have observed the transit-center was 51%, for the entire ingress it was 31% and for the egress 66%. The probability to have observed *either* in- or egress was 76%. Since no relevant brightness changes were found on any contaminator, we concluded therefore that the transit occurs likely on the target, but that a probability of $< 25\%$ remains that an eclipse event on a contaminant was missed due to the timing error. Estimations of this kind are now routinely performed for photometric follow-up observations and have led to a significant improvement in the reliability to detect, or to reject the presence of false-alarm sources.

## References

Aigrain, S. et al. 2009, A&A 506, 425
Deeg, H.J. et al. 2009, A&A 506, 343
Doyle, L.R. & Deeg, H.J. 2004, IAUS 213, 80
Jenkins, J.M. et al. 2010, ApJ 713, 120